\documentclass[12pt]{iopart}
\usepackage{iopams}
  \usepackage{amssymb}
  \usepackage{graphicx}

\catcode`@=11
\newcommand{\overset}[2]{\binrel@{#2}%
  \binrel@@{\mathop{\kern\z@#2}\limits^{#1}}}
\newcommand{\underset}[2]{\binrel@{#2}%
  \binrel@@{\mathop{\kern\z@#2}\limits_{#1}}}
\catcode`@=12

\newcounter{storeeqn}
\renewcommand{\theequation}{\arabic{section}.\arabic{equation}}

\newfont{\myeu}{eurm10 at 12 pt}
\newcommand{\half}{{\textstyle\frac{1}{2}}}
\newcommand{\halfs}{{\scriptstyle\frac{1}{2}}}
\def\sfactor#1#2{\raisebox{-2pt}{$\Bigg[$}{#1\atop#2}
 \raisebox{-2pt}{$\Bigg]$}}

\def\vp{^{\vphantom{x}}}
\def\vpt{_{\vphantom{'}}}
\def\vP{\vphantom{Q}P}

\def\bI{{\bar{\mathcal I}}}
\def\bA{{\bar A}}
\def\bB{{\bar B}}
\def\bC{{\bar C}}
\def\bD{{\bar D}}
\def\bK{{\bar K}}
\def\Bsf{\mbox{\boldmath$\mathsf B$}}
\def\Ysf{\mbox{\boldmath$\mathsf Y$}}
\def\Isf{\mbox{\boldmath$\mathsf 1$}}

\begin{document}
\title [Order Parameter in Chiral Potts Model]%
{Spontaneous Magnetization of the Integrable Chiral Potts Model}
\author{Helen Au-Yang$^{1,2}$ and Jacques H H Perk$^{1,2}$%
\footnote{Supported in part by the National Science Foundation
under grant PHY-07-58139 and by the Australian Research Council
under Project ID: LX0989627 and DP1096713.}}
\address{$^1$ Department of Physics, Oklahoma State University,
145 Physical Sciences, Stillwater, OK 74078-3072, USA%
\footnote{Permanent address.}\\
$^2$ Centre for Mathematics and its Applications \&\
Department of Theoretical Physics,
Australian National University, Canberra, ACT 2600, Australia}
\ead{\mailto{perk@okstate.edu}, \mailto{helenperk@yahoo.com}}

\begin{abstract}
We show how $Z$-invariance in the chiral Potts model provides
a strategy to calculate the pair correlation in the general
integrable chiral Potts model using only the superintegrable
eigenvectors. When the distance between the two spins in the
correlation function becomes infinite it becomes the square of
the order parameter. In this way, we show that the spontaneous
magnetization can be expressed in terms of the inner products of
the eigenvectors of the $N$ asymptotically degenerate maximum
eigenvalues. Using our previous results on these eigenvectors,
we are able to obtain the order parameter as a sum almost identical
to the one given by Baxter. This gives the known spontaneous
magnetization of the chiral Potts model by an entirely different
approach.
\end{abstract}

\section{Introduction\label{sec1}}
In 1988 Albertini {\it et al.}\ \cite{AMPT89} conjectured a simple
formula for the spontaneous magnetization ${\mathcal M}_r$ of the chiral
Potts model. It took many years to find a proof of this conjecture, as
the usual corner transfer matrix technique \cite{Baxterbook} could
not be used, because the variables on the rapidity lines of the chiral
Potts model \cite{AMPTY,BPAuY88} live on a higher genus curve and thus
do not satisfy the typical difference property. This conjecture was
finally proven by Baxter in 2005 \cite{RJB2005a,RJB2005b} using
functional equations and the ``broken rapidity line" technique of
Jimbo {\it et al.}\ \cite{JMN93}, invoking two rather mild analyticity
assumptions of the type commonly used in the field of Yang--Baxter
integrable models. Most recently, in a series of papers \cite{BaxterI,
BaxterII,BaxterIII,BaxterIV,BaxterV,Iorgov}, an
algebraic (Ising-like) way of obtaining ${\mathcal M}_r$ has been given,
providing more insight into the algebraic structure.

It now looks very plausible that the pair correlation function is also
calculable. In this paper we shall outline a strategy to attack this
problem. As a first application we shall derive the order parameter in
a new way.

\subsection{ Baxter's $Z$-invariance for correlation functions%
\label{sec1.1}}
\begin{figure}[tbh]
\begin{center}
\includegraphics[width=0.45\hsize]{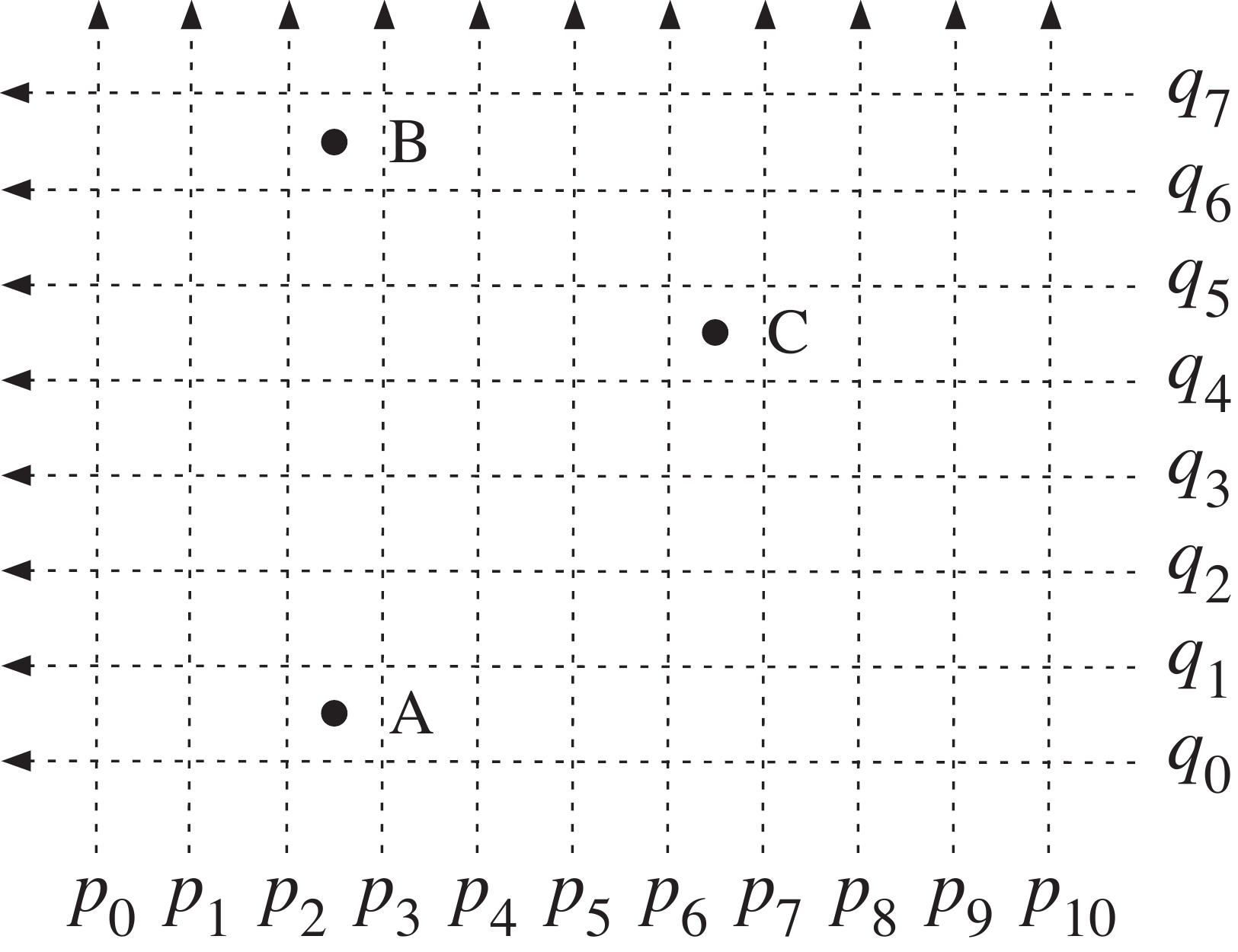}
\end{center}
\caption{\footnotesize Horizontal and vertical rapidity lines and
pair correlation functions
$\langle s_{\rm A}^{\,n}s_{\rm B}^{-n}\rangle=
g_6^{(n)}(q_1,q_2,q_3,q_4,q_5,q_6)$,
$\langle s_{\rm A}^{\,n}s_{\rm C}^{-n}\rangle=
g_8^{(n)}(q_1,q_2,q_3,q_4,p_3,p_4,p_5,p_6)$ and
$\langle s_{\rm C}^{\,n}s_{\rm B}^{-n}\rangle=
g_6^{(n)}(q_5,q_6,{\rm R}p_3,{\rm R}p_4,{\rm R}p_5,{\rm R}p_6)$.}
\label{corre}
\end{figure}


The integrable chiral Potts model \cite{BPAuY88} is $Z$-invariant by
definition---its transfer matrices commute with one another and one
can permute them without changing the partition function $Z$.
Furthermore, $Z$-invariance
also implies that in the thermodynamic limit the pair correlation
function $\langle s_{\rm A}^{\,n}s_{\rm B}^{-n}\rangle$ in the bulk is
given by a set of universal functions $g_{2m_{\vphantom{.}}}^{(n)}$,
which can only depend on the $2m$ rapidities that pass in the same
direction between the two spins at A and B \cite{BaxZI,APZI}.
This may require us to flip the direction of some rapidities by the
automorphism $q\to{\rm R}q$, given by $x_q\to y_q$, $y_q\to\omega x_q$,
to make all the $2m$ lines pointing in the same direction. Each spin
$s$ can take the $N$ values $s=\omega^{\sigma}$, with
$\sigma=0,\ldots,N\,-\,1$, $\omega\equiv\e^{2\pi\mathrm{i}/N}$.
Shown in figure \ref{corre} are three examples with universal
functions $g_6^{\,(n)}$ and $g_8^{\,(n)}$.

The $Z$-invariance property means that we can calculate the
correlation functions of a general integrable chiral Potts model
on an infinite lattice from special correlations in the much simpler
superintegrable case \cite{Baxter1988,Baxter1989,vonGehlen1985,JHHP89}.
We can take an infinite square lattice with special vertical rapidities
$p$ and $p'$ alternatingly, but with more general horizontal rapidities
$q_1,q_2,\cdots,q_{2m}$. Choose two spins within faces in the same
vertical column. As such a pair correlation function only depends on
modulus $k$ and the rapidities passing between the two spins, it is
independent of $p$ and $p'$, see also the example shown in figure
\ref{corre} as $\langle s_{A}^ns_{B}^{-n}\rangle$.
\subsection{Superintegrable chiral Potts model\label{sec1.2}}
\medskip
If we assume alternating vertical rapidities $x_{p'}=y_p$, $y_{p'}=x_p$,
$\mu_{p'}=1/\mu_p$, the model becomes ``superintegrable"
\cite{Baxter1988,Baxter1989,vonGehlen1985,JHHP89}, as it has Ising-like
properties. We shall consider here only the case with $p=p'$, so that
$\mu_p=1$. This case both obeys Yang--Baxter integrability
\cite{BPAuY88} and has an underlying Onsager algebra \cite{JHHP89},
whence the name superintegrable \cite{AMPT89}. Its spin chain
Hamiltonian was first constructed by von Gehlen and Rittenberg
\cite{vonGehlen1985}.

To calculate pair correlations of the superintegrable spin chain
in a ground state, we only need correlations in a horizontal row
of figure \ref{corre}, with uniform vertical rapidities satisfying
$x_{p}=y_p$. This pair correlation is the expectation value of
the spin pair in that ground state. From our previous work
\cite{AuYangPerk2008,AuYangPerk2009,AuYangPerk2010a,AuYangPerk2010b}
we know the $N$ ground state eigenvectors in the commensurate phase
that are needed.

Several partial results on the pair correlations in the superintegrable
quantum spin chain already exist, including some results for the leading
asymptotic long separation behavior from finite-size calculations and
conformal field theory \cite{AlbMcC1991,Cardy1993}, a few terms in
series expansions in the coupling constant
\cite{HanHon1994,Hon1994,Hon1995,HonvGe1995,McCoyOrrick1996},
results using the density matrix renormalization group technique
\cite{vGeRit2001} and a few exact results for very small chains
\cite{FabMcC2010}.
Using the exact knowledge of the eigenvectors much more can be done for
the superintegrable spin chain, but in this paper we shall rather go in
the vertical direction in figure \ref{corre}, as this will allow us to
describe what is needed for the more general integrable chiral Potts
model and to express the order parameters in terms of inner products
of ground state vectors.

\subsection{Correlation functions and order parameters\label{sec1.3}}
\setcounter{equation}{\value{storeeqn}}
 \renewcommand{\theequation}{\thesection.\arabic{equation}}
First we define the row-to-row transfer matrices with periodic boundary
conditions in the usual way \cite{Baxterbook}, {\it i.e.}
\begin{eqnarray}
T_q=T(x_q,y_q)_{\sigma \sigma'}=\prod_{J=1}^{L}
W_{p q}(\sigma^{\phantom{'}}_J -\sigma'_J)
{\overline W}_{p q}(\sigma^{\phantom{'}}_{J+ 1} -\sigma'_{J}),
\nonumber\\
{\hat T}_r={\hat T}(x_r,y_r)_{\sigma' \sigma''}=
\prod_{J=1}^{L}{\overline W}_{pr}(\sigma'_J -\sigma''_J)
W_{pr}(\sigma'_J -\sigma''_{J+ 1}),
\end{eqnarray}
where the chiral Potts Boltzmann weights depend on spin differences
modulo $N$.

Next, we define the vertical pair correlation functions, between spins
$\sigma_1$ and $\bar\sigma_1$ in the first column separated by $2\ell$
horizontal rapidities, as
\begin{eqnarray}
\fl g^{(r)}_{2\ell}(k;q_1,\cdots,q_{2\ell})=
{1\over Z}\sum_{\{\sigma\}}
\omega^{r(\sigma_1-{\bar\sigma}_{1})}
\prod_{{\rm all}\;{\rm bonds}}W_{pq}(\sigma-\sigma')
{\overline W}_{pq}(\sigma-\sigma''),
\end{eqnarray}
where $r=1,\ldots,N-1$. In the Ising case $N=2$ we only have $r=1$.
Assuming periodic boundary conditions in both directions, we can write
\begin{equation}
\fl g^{(r)}_{2\ell}(k;q_1,\cdots,q_{2\ell})=
{1\over Z}\mathop{\rm Tr}_{\{\sigma\}}
\bigg[{\bf \hat Z}_1^{\,r}\bigg(\prod_{j=1}^{\ell}
T_{q_{2j-1}}{\hat T}_{q_{2j}}\!\bigg)
{\bf \hat Z}_1^{\dagger r}\bigg(\prod_{j=\ell+1}^{M}
T_{q_{2j-1}}{\hat T}_{q_{2j}}\!\bigg)\bigg],
\end{equation}
with $\displaystyle
Z=\mathop{\rm Tr}_{\{\sigma\}}
\bigg(\prod_{j=1}^{M}
T_{q_{2j-1}}{\hat T}_{q_{2j}}\!\bigg)$. Once the thermodynamic limit
$L,M\to\infty$ is taken, this should give the most general pair
correlations in integrable chiral Potts models on infinite
$Z$-invariant lattices.

As in \cite{BaxterII}, the operators $\bf\hat Z$ and $\bf\hat X$ are
defined as ${\bf \hat Z}|\sigma\rangle=\omega^\sigma|\sigma\rangle$
and ${\bf \hat X}|\sigma\rangle=|\sigma\!+\!1\rangle$ and $L$ pairs
of copies ${\bf \hat Z}_j$ and ${\bf \hat X}_j$, acting in the $j$th
column, ($j=1,\ldots,L$), are introduced. They are different from
${\bf Z}_j$ and ${\bf X}_j$ of  \cite{AuYangPerk2010b} which act on
the edge variables  $n_j=\sigma_j-\sigma_{j+1}$, instead of the
spins $\sigma$. The transfer matrices are invariant under the spin
shift operator
\begin{equation}
{\mathcal X}\equiv\prod_{j=1}^L{\bf \hat X}_j, \qquad
{\mathcal X}|\sigma_1,\ldots,\sigma_L\rangle=
|\sigma_1\!+\!1,\ldots,\sigma_L\!+\!1\rangle.
\end{equation}
In fact, the transfer matrix elements only depend on differences
$n_j=\sigma_j\!-\!\sigma_{j+1}$ in both rows and
$m\equiv\sigma_1\!-\!\sigma'_1$ in the first column, {\it i.e.}
\begin{equation}
\langle\,\sigma'_1,\{n'_j\}|\, T\,|\sigma\vp_1,\{n_j\}\rangle=
T(\sigma\vp_1\!-\!\sigma'_1,\{n'_j\},\{n\vp_j\})=
T(m,\{n'_j\},\{n\vp_j\}),
\end{equation}
writing
\begin{equation}
|\sigma_1,\ldots,\sigma_L\rangle\equiv|\{\sigma_j\}\rangle\equiv
|\sigma_1,\{n_j\}\rangle,\quad n_j=\sigma_j\!-\!\sigma_{j+1},
\quad \sum_{j=1}^L n_j=0.
\end{equation}
We define a new (Fourier-transformed) basis \cite{AuYangPerk2008}
\begin{equation}
|Q\,;\{n_j\}\rangle\equiv N^{-1/2}\sum_{\sigma_1=0}^{N-1}
\omega^{-Q\sigma_1}|\sigma_1,\{n_j\}\rangle.
\end{equation}
These are eigenvectors of the spin shift operator, as
\begin{equation}
\fl{\mathcal X}\,|\sigma_1,\{n_j\}\rangle=|\sigma_1\!+\!1,\{n_j\}\rangle,
\quad\hbox{implies}\quad
{\mathcal X}\,|Q\,;\{n_j\}\rangle=\omega^Q|Q\,;\{n_j\}\rangle.
\end{equation}
In this new basis, the transfer matrix elements become
\begin{equation}
\fl\langle\,Q\,;\{n'_j\}|\, T\,|Q\,;\{n\vp_j\}\rangle=
{1\over N}\sum_{\sigma'_1=0}^{N-1}\sum_{\sigma\vp_1=0}^{N-1}
\omega^{Q(\sigma'_1-\sigma\vp_1)}
\langle\,\sigma'_1,\{n'_j\}|\, T\,|\sigma\vp_1,\{n\vp_j\}\rangle,
\end{equation}
or
\begin{equation}
\fl\langle\,Q\,;\{n'_j\}|\, T\,|Q\,;\{n\vp_j\}\rangle=
\sum_{m=0}^{N-1}\omega^{-mQ}\,T(m,\{n'_j\},\{n\vp_j\})\equiv
T_Q(\{n'_j\},\{n\vp_j\}).
\end{equation}
Thus we find that in the spin-shift $Q$ sector the transfer
matrices only depend on bond (link, or edge) variables.

The inverse relation is
\begin{equation}
\langle\{\sigma'_j\}|\, T\,|\{\sigma\vp_j\}\rangle=
{1\over N}\sum_{Q=0}^{N-1}\omega^{Q(\sigma\vp_1-\sigma'_1)}
T_Q(\{n'_j\},\{n\vp_j\}).
\end{equation}
Matrix products can also be rewritten in the $Q$ basis:
\begin{eqnarray}
\fl\langle\{\sigma'_j\}|\,T\hat T\,|\{\sigma\vp_j\}\rangle&=&
\sum_{\sigma''_1=0}^{N-1}\sum_{\{n''\}}
\langle\{\sigma'_j\}|\,T\,|\{\sigma''_j\}\rangle
\langle\{\sigma''_j\}|\,\hat T\,|\{\sigma\vp_j\}\rangle
\nonumber\\
&=&{1\over N^2}\sum_{\sigma''_1=0}^{N-1}\sum_{\{n''\}}
\sum_{Q=0}^{N-1}\sum_{Q'=0}^{N-1}
\omega^{Q(\sigma''_1-\sigma'_1)}T_Q(\{n'_j\},\{n''_j\})
\nonumber\\
&&\qquad\times\,\omega^{Q'(\sigma\vp_1-\sigma''_1)}
\hat T_{Q'}(\{n''_j\},\{n\vp_j\})
\nonumber\\
&=&{1\over N}\sum_{\{n''\}}
\sum_{Q=0}^{N-1}\sum_{Q'=0}^{N-1}
\omega^{Q(\sigma\vp_1-\sigma'_1)}\delta_{Q,Q'}T_Q(\{n'_j\},\{n''_j\})
\hat T_{Q'}(\{n''_j\},\{n\vp_j\})
\nonumber\\
&=&{1\over N}\sum_{\{n''\}}\sum_{Q=0}^{N-1}
\omega^{Q(\sigma\vp_1-\sigma'_1)}T_Q(\{n'_j\},\{n''_j\})
\hat T_{Q}(\{n''_j\},\{n_j\})
\nonumber\\
&=&{1\over N}\sum_{Q=0}^{N-1}\omega^{Q(\sigma\vp_1-\sigma'_1)}
\langle\{n'_j\}|\,T_Q
\hat T_{Q}\,|\{n\vp_j\}\rangle.
\end{eqnarray}
The pair correlation function can be worked out similarly.
In the special case of equal horizontal rapidities ($q_{\ell}\equiv q$)
and using $\langle\{\sigma\}|{\bf\hat Z}_1^{\,r}|\{\sigma\}\rangle=
\omega^{r\sigma_1}$, we find
\begin{eqnarray}
\fl g^{(r)}_{2\ell}(k;q,\cdots,q)
=\frac 1{ Z}\mathop{\rm Tr}_{\{\sigma\}}
\bigg[{\bf\hat Z}_1^{\,r}\bigg(\prod_{j=1}^{\ell}
T_{q_{2j-1}}{\hat T}_{q_{2j}}\!\bigg)
{\bf\hat Z}_1^{\dagger r}\bigg(\prod_{j=\ell+1}^{M}
T_{q_{2j-1}}{\hat T}_{q_{2j}}\!\bigg)\bigg]
\nonumber\\
\fl=\frac 1{ Z}\sum_{Q=0}^{N-1}\mathop{\rm Tr}_{\{n_j\}}
\Bigl\{\Bigl[T_P(x_q,y_q){\hat T}_P(x_q,y_q)\Bigr]^{\ell}
\Bigl[T_Q(x_q,y_q){\hat T}_Q(x_q,y_q)\Bigr]^{M-\ell}\Bigr\},
\end{eqnarray}
with $P\equiv Q-r$ mod $N$. Let the eigenvectors of the transfer
matrices be given by
\begin{equation}
T_Q(x_q,y_q){\hat T}_Q(x_q,y_q)|{\mathcal Y}_j^Q\rangle=
\big(\Delta_j^Q\big)^2|{\mathcal Y}_j^Q\rangle,\qquad
\langle{\mathcal Y}_i^Q|{\mathcal Y}_j^Q\rangle=\delta_{i,j},
\end{equation}
where $\Delta_{j}^Q$ denotes the $j$th eigenvalue,
and let $\Delta_{\mathrm{max}}^Q$ be the maximum eigenvalue of
the transfer matrix $T_Q$.
Then, in the limit of an infinite number $2M$ of rows, the partition
function becomes
\begin{equation}
Z=\sum_{Q=0}^{N-1}(\Delta^Q_{\mathrm{max}})^{2M}\to
N(\Delta_{\mathrm{max}}^0)^{2M},\quad\hbox{as}\quad L,M\to \infty,
\end{equation}
as the $\Delta_{\mathrm{max}}^Q$ for $0\le Q\le N\!-\!1$ are
asymptotically degenerate as $L\to\infty$. Therefore,
\begin{equation}
g^{(r)}_{2\ell}(k;q,\cdots,q)={1\over N}\sum_{Q=0}^{N-1}
\sum_{j=1}^J\Biggl[{{\Delta^P_j}\over{\Delta^0_{\mathrm{max}}}}
\Biggr]^{2\ell}
\langle{\mathcal Y}_{\mathrm{max}}^Q|{\mathcal Y}_j^P\rangle
\langle{\mathcal Y}_j^P|{\mathcal Y}^Q_{\mathrm{max}}\rangle,
\label{gr2l}\end{equation}
where $J=N^{L-1}$, $L\to\infty$.

In the limit $\ell\to\infty$, the pair correlation becomes
the product of order parameters,
\begin{equation}
\fl\langle\omega^{r\sigma_1}\rangle\langle\omega^{-r\sigma_1}\rangle
={1\over N}\sum_{Q=0}^{N-1}
\langle{\mathcal Y}_{\mathrm{max}}^Q|
{\mathcal Y}_{\mathrm{max}}^P\rangle
\langle{\mathcal Y}_{\mathrm{max}}^P|
{\mathcal Y}^Q_{\mathrm{max}}\rangle,
\quad P\equiv Q-r \mbox{ mod $N$}.
\label{order}\end{equation}
For the Ising case ($N\!=\!2$) this reduces to the simple formula for
the spontaneous magnetization $m_{\mathrm{sp}}=
|\langle{\mathcal Y}_{\mathrm{max}}^0|
{\mathcal Y}_{\mathrm{max}}^1\rangle|$
in the ordered phase.

\subsection{Baxter's approach\label{sec1.4}}
Recognizing that the $N$-state superintegrable model resembles the
2d Ising model \cite{AMPT89,Baxter1989,AuYangPerk2009,%
AuYangPerk2010b,Davies1990,NiDe1,NiDe2}, and that the two parts of
its Hamiltonian generate the Onsager algebra
\cite{vonGehlen1985,Davies1990}, Baxter concluded that there must an
algebraic route to calculate the order parameter ${\mathcal M}_r$.
He started \cite{BaxterI} with a new algebraic approach to calculate
${\mathcal M}_r$ for the Ising model, representing it first as a square
root of an $L$ by $L$ determinant, which he then reduced to an $m$ by
$m$ determinant \cite{BaxterI}, where $m\le L/2$. Having these new
Ising results Baxter tried to guess their extensions to the $N$-state
superintegrable model. Thus he conjectured \cite{BaxterII,BaxterIII}
that the special combination
$D_{PQ}={\mathcal M}_r({\mathcal Z}_P{\mathcal Z}_Q)^{1/2}$,
with ${\mathcal Z}_Q$ denoting the partition function with fixed
boundary condition of type $Q$, is both a determinant and also
given by the sum
\begin{equation}
D_{PQ}=\sum_{s\vp}\sum_{s'}
\,y_1^{s_1}y_2^{s_2}\cdots y_m^{s_m}
\left(\frac{A_{s,s'}B_{s,s'}}{C_s D_{s'}}\right)
{y'_1}^{s'_1}{y'_2}^{s'_2}\cdots
{y'_{m'}}^{\!\!s'{\!\!\raisebox{-2pt}{$\scriptscriptstyle{m'}$}}}.
\label{LHS}\end{equation}
Here $m=m_P$ and $m'=m_Q$, with $m_Q=\lfloor(N-1)L/N-Q/N\rfloor$
and $m_P$ given similarly,
\begin{equation}
s=\{s_1,s_2,\ldots,s_m\},\quad s'=\{s'_1,s'_2,\ldots,s'_{m'}\},
\quad s_i\vp,s'_i=0,1,
\label{ssp}\end{equation}
and
\begin{eqnarray}
A_{s,s'}=\prod_{i\,\in\,W}\prod_{j\,\in V'} (c_i\vp-c'_j),&\quad&
B_{s,s'}=\prod_{i\,\in\,V }\prod_{j\,\in W'}(c_i\vp-c'_j),
\nonumber\\
C_{s}\,=\,\prod_{i\,\in\,W }\prod_{j\,\in V } (c_j\vp-c_i\vp),&&
D_{s'}\,=\,\prod_{\,i\,\in V'}\prod_{j\,\in W'}(c'_j-c'_i),
\label{defADT}\end{eqnarray}
in which for a given set $s$, $V$ denotes the set of integers $i$ such
that $s_i= 0$ and $W$ the set such that $s_i = 1$, while $V'$, $W'$
are defined similarly for the set $s'$. In (\ref{LHS}) the number of
elements in $W$ equals the number in $W'$. The variables $c_i$ are
related to the roots $w_i$ of $P_P(w)$ and $c'_i$ to the roots $w'_i$
of $P_Q(w)$, with both polynomials defined by
\begin{equation}
P_Q(t^N)=t^{-Q}\sum_{n=0}^{N-1}\omega^{-nQ}
\left[\frac{(1-t^N)}{(1-t\omega^n)}\right]^L
\label{dfp}\end{equation}
and the $c_i\vp$ and $c'_i$ by the relation $c=(w+1)/(w-1)$.
This form (\ref{LHS}) was recently proven by Iorgov {\it et al.}\
\cite{Iorgov}. Finally, in \cite{BaxterV}, Baxter showed that this sum
$D_{PQ}$ is the conjectured determinant, which he had already
evaluated in the thermodynamic limit in \cite{BaxterIV}. Thus he
completed an algebraic proof of the order parameter.

\subsection{Approach of Iorgov et al.\label{sec1.5}}

Very recently Iorgov {\it et al.}\ \cite{Iorgov} derived matrix
elements of the spin operator in the finite superintegrable chiral
Potts quantum chain in factorized form. They also gave another
method to derive the order parameters in the thermodynamic limit,
performing the sum without going to the determinant formula of
Baxter.

\section{Our Approach\label{sec2}}
\setcounter{equation}{\value{storeeqn}}
 \renewcommand{\theequation}{\thesection.\arabic{equation}}

In our previous papers \cite{AuYangPerk2009,AuYangPerk2010b} we have
given the $2^{m_Q}$ eigenvector pairs in the ground state sectors of
the transfer matrices $T_Q$ and $\hat T_Q$. Using the notation of
Baxter \cite{BaxterIII,BaxterIV}, but without going to Baxter's
reduced representation of the vector space, we label these eigenvectors
$|{\bf \mathcal Y}^Q_{s'}\rangle$ and $|{\bf \mathcal X}^Q_{s'}\rangle$,
with $s'$ chosen as in (\ref{ssp}). In fact, the eigenvectors
are given in \cite{AuYangPerk2009,AuYangPerk2010b} as
\begin{equation}
|{\bf \mathcal X}^Q_{s'}\rangle=\prod_{j=1}^{m_Q}{\bf \mathcal R}_{j,Q}
\prod_{i\in W'_n}{\bf E}_{i,Q}^{+}|\Omega\rangle,\quad
|{\bf \mathcal Y}^Q_{s'}\rangle=\prod_{j=1}^{m_Q}{\bf \mathcal S}_{j,Q}
\prod_{i\in W'_n}{\bf E}_{i,Q}^{+}|\Omega\rangle,
\label{eigenvectors}\end{equation}
where $W'_n$ is a subset of $\{1,2,\cdots,m_Q\}$ containing $n$
integers. They satisfy the eigenvalue equations
\begin{equation}
{\mathcal T}_Q(x_q,y_q)|{\bf \mathcal X}^Q_{s'}\rangle=
\Delta^Q_{s'}|{\bf \mathcal Y}^Q_{s'}\rangle,\quad
{\hat{\mathcal T}}_Q(y_q,x_q)|{\bf \mathcal Y}^Q_{s'}\rangle=
\Delta^Q_{s'}|{\bf \mathcal X}^Q_{s'}\rangle.
\end{equation}
Particularly, we shall use the ground state eigenvectors
\begin{equation}
|{\mathcal Y}_{\mathrm{max}}^P\rangle=
|{\mathcal Y}_{\emptyset}^P\rangle=
\prod_{j=1}^{m_P}{\bf \mathcal S}_{j,P}|\Omega\rangle,\quad
\langle{\mathcal Y}_{\mathrm{max}}^P|=
\langle{\mathcal Y}_{\emptyset}^P|=
\langle\Omega|\prod_{j=1}^{m_P}{\bf \mathcal S}^{-1}_{j,P}\,,
\label{PP}\end{equation}
for which $W_{\emptyset}=\emptyset$ is the empty set. The ``rotation"
${\bf\mathcal S}=\prod_j{\bf\mathcal S}_j$ is given by (IV.124) or
(IV.150)\footnote{Equations in our papers
\cite{AuYangPerk2008,AuYangPerk2009,AuYangPerk2010a,AuYangPerk2010b}
are denoted here by respectively prefacing I, II, III, or IV to
the equation number.} with
\begin{eqnarray}
{\bf \mathcal S}_{j,P}=\half(s^{j,P}_{11}+s^{j,P}_{22}){\bf 1}+
\half(s^{j,P}_{11}-s^{j,P}_{22}){\bf H}_{j,P}+
s^{j,P}_{12}{\bf E}^+_{j,P}+s^{j,P}_{21}{\bf E}^-_{j,P}\,,
\label{Sj}\\
{\bf \mathcal S}^{-1}_{j,P}=\half(s^{j,P}_{22}+s^{j,P}_{11}){\bf 1}+
\half(s^{j,P}_{22}-s^{j,P}_{11}){\bf H}_{j,P}-
s^{j,P}_{12}{\bf E}^+_{j,P}-s^{j,P}_{21}{\bf E}^-_{j,P}\,,
\label{S1j}\end{eqnarray}
where ${\bf E}^\pm_{j,P}$ and ${\bf H}_{j,P}$ are generators of
${\mathfrak{sl}}_2$ algebras. Furthermore, we have shown in
\cite{AuYangPerk2009,AuYangPerk2010b} that for the state
$|\Omega\rangle\equiv|\{n_j\!=\!0\}\rangle$ we have
${\bf E}^{-}_{m,Q}|\Omega\rangle\!=\!0$,
${\bf H}_{m,Q}|\Omega\rangle\!=\!-|\Omega\rangle$,
$\langle\Omega|{\bf E}_{\ell,Q}^{+}\!=\!0$,
$\langle\Omega|{\bf H}_{\ell,Q}\!=\!-\langle \Omega|$ and
\begin{eqnarray}
\langle\Omega|{\bf E}_{\ell,Q}^{-}=
-\frac{\beta^Q_{\ell,0}}{\Lambda^Q_{0}}
\sum_{ {\{0\le n_j\le N-1\}}\atop{n_1+\cdots+n_L=N}}\langle\{n_j\}|\,
\omega^{\sum_{j} j n_j}\bar{G}^{\vp}_Q(\{n_j\},z^{\vp}_{\ell,Q}),
\nonumber\\
{\bf E}_{k,Q}^{+}|\Omega\rangle=
\frac{\beta^Q_{k,0}z^{\vp}_{k,Q}}{\Lambda^Q_{0}}
\sum_{ {\{0\le n_j\le N-1\}}\atop{n_1+\cdots+n_L=N}}
\omega^{\!-\!\sum_{j} j n_j}
{G}^{\vp}_Q(\{n_j\},z^{\vp}_{k,Q})|\{n_j\}\rangle,
\label{eo}\end{eqnarray}
see (IV.66) and (IV.67).
The polynomials here are given in (III.16) as
\begin{eqnarray}
\fl\qquad{\bar G}_Q(\{n_j\},z)
=\!\sum_{n=0}^{m'-1}\bK_{nN+Q}(\{n_j\})z^n,\quad
G_P(\{n_j\},z)
=\!\sum_{n=0}^{m-1}K_{nN+P}(\{n_j\})z^n,
\label{GbG}\end{eqnarray}
with $m\equiv m_P$, $m'\equiv m_Q$ and
coefficients given by (III.7) and (III.8) as the sums
\begin{eqnarray}
&&K_\ell(\{n_j\})=
\sum_{ {\{0\le n'_j\le N-1\}}\atop{n'_1+\cdots+n'_L=\ell} }
\prod_{j=1}^L \sfactor{n\vp_j+n'_j}{n'_j}
\omega^{n'_j N\vp_j},\quad N_j=\sum_{\ell=1}^{j-1}n_\ell\,,
\nonumber\\
&&\bK_\ell(\{n_j\})=
\sum_{{\{0\le n'_j\le N-1\}}\atop{n'_1+\cdots+n'_L=\ell}}
\prod_{j=1}^L\sfactor{n\vp_j+n'_j}{n'_j}
\omega^{n'_j{\bar N}\vp_j},\quad {\bar N}_j
=\sum_{\ell=j+1}^{L}n_\ell\,.
\label{KbK}\end{eqnarray}

\subsection{Form factors\label{sec2.1}}
From (\ref{PP}), (\ref{Sj}) and (\ref{S1j}) we obtain
\begin{eqnarray}
\fl\langle{\mathcal Y}_{\mathrm{max}}^Q|
{\mathcal Y}^P_{\mathrm{max}}\rangle=
\langle\Omega|
\prod_{j=1}^{m_Q}(s^{j,Q}_{11}{\bf 1}-s^{j,Q}_{21}{\bf E}^-_{j,Q})
\prod_{j=1}^{m_P}(s^{j,P}_{22}{\bf 1}+s^{j,P}_{12}{\bf E}^+_{j,P})
|\Omega\rangle
\nonumber\\
=\prod_{j=1}^{m'}{s'}^{j}_{11}\prod_{j=1}^{m}s^{j}_{22}\,
\langle\Omega|\prod_{j=1}^{m'} (1+u'_j{\bf E}^-_{j,Q})
 \prod_{j=1}^{m} (1+u\vp_{j}{\bf E}^+_{j,P})|\Omega\rangle,
\label{QyP}\end{eqnarray}
where $m=m_P$, $m'=m_Q$, and
\begin{equation}
u\vp_j={s_{12}^j\over s_{22}^j}=\frac{s^{j,P}_{12}}{s^{j,P}_{22}}=
\frac{T^*_{12}}{T^*_{22}},\quad
u'_j=-{s'^j_{21}\over s'^j_{11}}=-\frac{s^{j,Q}_{21}}{s^{j,Q}_{11}}=
-\frac{T'_{21}}{T'_{11}},
\label{yj}\end{equation}
see (II.C.2), (II.C.10) and (IV.152).

Next we expand the products in (\ref{QyP}). Since
\begin{eqnarray}
\prod_{s=1}^k {\bf E}^+_{\ell_s,P}\;
|\Omega\rangle\in{\textstyle\bigoplus}\Big\{|\{n_j\}\rangle,\;\
\sum_j n_j=k N\Big\},
\nonumber\\
\langle\Omega| \prod_{s=1}^l {\bf E}^-_{\ell_s,P}
\in{\textstyle\bigoplus}\Big\{\langle\{n'_j\}|,\;\
\sum_j n'_j=l N\Big\},
\end{eqnarray}
we find that the only non-vanishing terms in this expansion are
those with equal numbers of creation and annihilation operators. 
Consequently,
\begin{eqnarray}
\fl\langle{\bf\mathcal Y}^Q_{\emptyset}
|{\bf\mathcal Y}^{\vP}_{\emptyset}\rangle=
{\mathcal C}\Bigg[ 1\!+\!\sum_{j=1}^{m'}
\sum_{\ell=1}^{m}u'_ju\vp_\ell\langle\Omega|
{\bf E}^-_{j,Q}{\bf E}^+_{\ell,P}|\Omega\rangle\!+\!\cdots
\nonumber\\
\fl+\mathop{\sum\cdots\sum}_{\scriptstyle 1\le j_1<\cdots< j_n\le m}\;
\mathop{\sum\cdots\sum}_{\scriptstyle 1\le\ell_1<\cdots<\ell_n\le m'}
(u\vp_{\ell_1}\!\cdots\!u\vp_{\ell_n})(u'_{j_1}\!\cdots\! u'_{j_{n}})
\langle\Omega|\prod_{s=1}^n {\bf E}^-_{j_s,Q}
\prod_{s=1}^n {\bf E}^+_{\ell_s,P}|\Omega\rangle
\nonumber\\
+\cdots+(u\vp_{1}\cdots u\vp_{m})(u'_{1}\cdots u'_{m'})
\langle\Omega|\prod_{j=1}^{m'}{\bf E}^-_{j,Q}
\prod_{\ell=1}^{m}{\bf E}^+_{\ell,P}|\Omega\rangle\Bigg],
\label{QP}\end{eqnarray}
with
\begin{equation}
{\mathcal C}\equiv\prod_{j=1}^{m'}{s'}^{j}_{\!\!11}
\prod_{j=1}^{m}s^{j}_{22}\,.
\label{QPC}\end{equation}
The last term in (\ref{QP}) is identically zero unless $m=m'$.
Let $\lambda_p=\mu^N_p=1$ in (III.149) and (II.C.5),
and denote $z\vp_j=z\vp_{j,P}$, $z'_j=z\vp_{j,Q}$,
$\theta\vp_j=\theta\vp_{j,P}$
and $\theta'_j=\theta\vp_{j,Q}$ as in Baxter's papers.
Then (\ref{yj}) becomes
\begin{eqnarray}
u\vp_j=\frac{1-k'}{\e^{2\theta_j}-k'},\quad
u'_j=-\frac{z'_j(1-k')}{\e^{2\theta'_j}-k'},
\label{yj2}\end{eqnarray}
compare (III.124) and (III.126) with $\lambda_p=1$.
The $\theta_j$ in these equations are not the same as those
in Baxter's papers \cite{BaxterII,BaxterIII}, but are related to
our $z_j$ by (II.C.5), {\it i.e.}
\begin{equation}
\e^{2\theta_j}+\e^{-2\theta_j}=k'+1/k'-(1-k')^2z_j/k'.
\label{eqet}\end{equation}
Solving for $\e^{2\theta_j}$ while using Baxter's notations
(BaxII.3.16) and (BaxII.2.18)\footnote{Equations in Baxter's papers
\cite{BaxterI,BaxterII,BaxterIII,BaxterIV,BaxterV}
are denoted here by respectively prefacing BaxI, BaxII, BaxIII,
BaxIV or BaxV to the equation number. Our transfer matrices
and $\mathbf{X}$ are transposes of the ones of Baxter, see footnote
to (IV.107). Therefore, to compare with Baxter, we must replace
$Q\to N-Q$ or $z_j\to1/z_j=w_j$.}
\begin{equation}
\lambda_j\equiv
\sqrt{1+k'^2+2k'\frac{1+z_j}{1-z_j}}\quad{\hbox{or}}\quad
z_j=\frac{\lambda_j^2-(1+k')^2}{\lambda_j^2-(1-k')^2},
\label{lz}\end{equation}
identifying $w_j\equiv1/z_j$ and
$c_j\equiv\cos(\theta^{\mathrm{B}}_j)=-(1+z_j)/(1-z_j)$,
(where we use $\theta^{\mathrm{B}}_j$
to denote $\theta_j$ in Baxter's paper), we have
\begin{equation}
{\rm e}^{2\theta_j}=\frac{\lambda\vp_j+1-k'}{\lambda\vp_j-1+k'}.
\label{et}\end{equation}
Consequently (\ref{yj2}) becomes
\begin{equation}
u_j=\frac{\lambda\vp_j-1+k'}{\lambda\vp_j+1+k'},\qquad
u'_j=-\frac{\lambda'_j-1-k'}{\lambda'_j+1-k'}.
\label{yj4}\end{equation}
where the second equation in (\ref{lz}) is used to get rid of the
factor $z'_j$. Comparing with (BaxIV.4.13) and (BaxIV.4.14), we
find $u'_j\to -y_j$ and $u_j\to -y'_j$. The minus signs cancel out
upon multiplication. We shall show later that this gives the correct
result, due to the identification $z_j=1/w_j$.

Adopting the notations of Baxter as shown in (\ref{ssp}) and
(\ref{defADT}), we
find $W_n=\{\ell_1,\cdots,\ell_n\}$ and $W'_n=\{j_1,\cdots,j_n\}$,
so that (\ref{QP}) becomes
\begin{eqnarray}
\langle{\bf \mathcal Y}^Q_{\emptyset}
|{\bf \mathcal Y}^{\vP}_{\emptyset}\rangle=&{\mathcal C}\,
\sum_{s\vp}\sum_{s'}\,u_1^{\;s\vp_1}u_2^{\;s\vp_2}\cdots u_m^{\;s\vp_m}
\nonumber\\
&\times\langle\Omega|\prod_{i\in W'_n}{\bf E}^-_{i,Q}
\prod_{j\in W_n}{\bf E}^+_{j,P}|\Omega\rangle
{u'_1}^{s'_1}{u'_2}^{s'_2}\cdots
{u'_{m'}}^{\!\!s'{\!\!\raisebox{-2pt}{$\scriptscriptstyle{m'}$}}}.
\label{yQyP}\end{eqnarray}
Similarly, for $P\leftrightarrow Q$, we have
\begin{eqnarray}
\langle{\bf\mathcal Y}^{\vP}_{\emptyset}
|{\bf\mathcal Y}^{Q}_{\emptyset}\rangle=
&{\hat{\mathcal C}}\,\sum_{s\vp}\sum_{s'}
{\hat u}_1^{\;s\vp_1}{\hat u}_2^{\;s\vp_2}\cdots{\hat u}_m^{\;s\vp_m}
\nonumber\\
&\times\langle\Omega|\prod_{j\in W_n}{\bf E}^-_{j,P}
\prod_{i\in W'_n}{\bf E}^+_{i,Q}|\Omega\rangle
{\hat u'}_1{}^{s'_1}{\hat u'}_2{}^{s'_2}\cdots{{\hat u'}_{m'}}
{}^{\!\!s'{\!\!\raisebox{-2pt}{$\scriptscriptstyle{m'}$}}},
\label{yPyQ}\end{eqnarray}
where, instead of (\ref{yj}) and (\ref{QPC}),
\begin{eqnarray}
{\hat u}'_j=s'^j_{12}/s'^j_{22},\quad
{\hat u}\vp_j=-s^j_{21}/s^j_{11},
\label{upj}\\
{\hat{\mathcal C}}=\prod_{j=1}^{m}{s}^{j}_{11}
\prod_{j=1}^{m'}{s'}^{j}_{\!\!22}.
\label{chc}\end{eqnarray}
It is easily seen from (\ref{yj}), (\ref{yj2}) and (\ref{upj}),
followed by (\ref{yj4}) and (\ref{lz}), that
\begin{equation}
{\hat u}\vp_j=-z\vp_j u\vp_j=
-\frac{\lambda\vp_j-1-k'}{\lambda\vp_j+1-k'},\qquad
{\hat u'}_j=-\frac{u'_j}{z'_j}=
\frac{\lambda'_j-1+k'}{\lambda'_j+1+k'}.
\label{ypj}\end{equation}
Finally from (II.C.6) and (\ref{et}), we have
\begin{equation}
s^{j}_{11}s^{j}_{22}=\frac{{\rm e}^{2\theta_j}-k'}{2\sinh2\theta_j}=
\frac{(\lambda_j+1)^2-k'^2}{4\lambda_j},
\label{ss}\end{equation}
which relates to the inverse of $Z_P$ in (BaxIV.3.7), as it should.

Looking at (\ref{yQyP}) we now need to evaluate the expectation value
in the state $|\Omega\rangle$ of the product of $n$ creation operators
${\bf E}^+_{j,P}$ and $n$ annihilation operators ${\bf E}^-_{j,Q}$,
which all operate on the edge variables.
\section{Proposition\label{sec3}}
\setcounter{equation}{\value{storeeqn}}
 \renewcommand{\theequation}{\thesection.\arabic{equation}}
Comparing (\ref{yQyP}) with the sum $D_{PQ}$ of Baxter
\cite{BaxterII,BaxterIII} given here in (\ref{LHS}), we propose
the following identity
\begin{eqnarray}
\langle\Omega|\prod_{j\in W_n'}{\bf E}^-_{j,Q}
\prod_{\ell\in W_n}{\bf E}^+_{\ell,P}|\Omega\rangle=
\frac {\bA_{s,s'}\bB_{s,s'}} {\bC_s \bD_{s'}}\propto
\frac {A_{s,s'}B_{s,s'}} {C_s D_{s'}},
\label{nEE}\end{eqnarray}
where $A$, $B$, $C$ and $D$ are defined in (\ref{defADT})
[or (BaxIII.3.44)], while
\begin{eqnarray}
\bA_{s,s'}=\prod_{i\in W_n}\prod_{j\in V_n'}(z\vp_i-z'_j),\quad
\bB_{s,s'}=\prod_{i\in W_n'}\prod_{j\in V_n}(z'_i-z\vp_j),
\nonumber\\
\bC_s=\prod_{i\in W_n}\prod_{j\in V_n}(z\vp_i-z\vp_j),\qquad
\bD_{s'}=\prod_{i\in W'_n}\prod_{j\in V'_n}(z'_i-z'_j).
\label{ABCD}\end{eqnarray}
The difference between $A$ and $D$ of Baxter and $\bA$ and $\bD$ here
is the replacement of $c=(z+1)/(z-1)$ by $z$. In $\bB$ and $\bC$,
we have also flipped the signs to make them more symmetric.

We shall first consider the simplest case, with $n=1$, then prove
(\ref{nEE}) by induction.

\subsection{Proof of (\ref{nEE}) for $n=1$\label{sec3.1}}
From (\ref{eo}), we find
\begin{eqnarray}
\fl\langle\Omega|{\bf E}_{j,Q}^{-}{\bf E}_{\ell,P}^{+}|\Omega\rangle
=-\frac{\beta^Q_{j,0}\beta^{\vphantom{Q}P}_{\ell,0}z\vp_{\ell,P}}
{\Lambda^Q_{0}\Lambda^{\vphantom{Q}P}_{0}}
\sum_{ {\{0\le n_i\le N-1\}}\atop{n_1+\cdots+n_L=N}}
\bar{G}\vp_Q(\{n_i\},z\vp_{j,Q})
G\vp_P(\{n_i\},z\vp_{\ell,P}).
\label{oeeo}\end{eqnarray}
Similar to (II.72) or (III.45), we introduce the polynomial
\begin{equation}
{\mbox{\myeu h}}^{Q,P}(z'_k,z)\equiv
\sum_{{\{0\le n_i\le N-1\}}\atop{n_1+\cdots+n_L=N}}
{\bar G}\vp_Q(\{n\vp_i\},z'_{k})G\vp_P(\{n_i\},z).
\label{dh}\end{equation}
Substituting (\ref{GbG}) into this equation, we find
\begin{eqnarray}
{\mbox{\myeu h}}^{Q,P}(z'_k,z)\equiv
\sum_{\ell=0}^{m'-1}\sum_{j=0}^{m-1}z'^\ell _{k}z\vpt^j
{\mathcal G}\vp_{\ell N+Q,jN+P},
\label{hqp1}\end{eqnarray}
where
\begin{equation}
{\mathcal G}_{\ell,j}=
\sum_{{\{0\le n_i\le N-1\}}\atop{n_1+\cdots+n_L=N}}
\bK_\ell(\{n_i\})K_j(\{n_i\}).
\label{coeff}\end{equation}
We can generalize the identity (III.37) or (III.44) to $P\ne Q$ cases.
For $P\ge Q$ we have
\begin{eqnarray}
{\mathcal G}_{\ell N+Q,jN+P}= {\mathcal G}_{j N+P,\ell N+Q}
\nonumber\\
=\sum_{n=0}^j\,[(j-n+1)\Lambda^{Q}_{n} \Lambda^{P}_{\ell+1+j-n}
-(n-\ell) \Lambda^{Q}_{\ell+1+j-n}\Lambda^{P}_{n}].
\label{Id}\end{eqnarray}
which becomes (III.37) for $P=Q$ after replacing $n\to j-n$.
The proof of (\ref{Id}) is much harder, however, and
in the Appendix we shall show why. We have been able to prove
the identity using a different approach, which is presented
in detail elsewhere \cite{AuYangPerk2011}. Due to the symmetry
${\mathcal G}_{\ell,k}={\mathcal G}_{k,\ell}$ given in (\ref{Id}),
we find
\begin{eqnarray}
{\mbox{\myeu h}}^{Q,P}(z'_k,z)=
\sum_{\ell=0}^{m'-1}\sum_{j=0}^{m-1}z'^\ell_{k}\, z_{\vp}^j
{\mathcal G}\vp_{jN+P,\ell N+Q}
={\mbox{\myeu h}}^{P,Q}(z,z'_k).
\label{hpqhqp}\end{eqnarray}
Inserting (\ref{Id}), and interchanging the order of summations
over $j$ and $n$, we rewrite (\ref{hqp1}) as
\begin{eqnarray}
\fl{\mbox{\myeu h}}^{Q,P}(z'_k,z)
=\sum_{\ell=0}^{m'-1}\sum_{n=0}^{m-1}\sum_{j=n}^{m-1}z'^\ell_{k}
z\vpt^j[(j-n+1)\Lambda^{Q}_{n} \Lambda^{P}_{\ell+1+j-n}
-(n-\ell) \Lambda^{Q}_{\ell+1+j-n}\Lambda^{P}_{n}]
\nonumber\\
=\sum_{\ell=0}^{m'-1}{\bigg(\frac {z'_{k}}z\bigg)}^{\!\ell}
\,\sum_{n=0}^{m-1}\sum_{i=\ell+1}^{m+\ell-n}
[(i-\ell)\Lambda^{Q}_{n} \Lambda^{P}_{i}
-(n-\ell) \Lambda^{Q}_{i}\Lambda^{P}_{n}]z^{n+i-1},
\label{hqp2}\end{eqnarray}
where the summation over $j$ is changed to $i=j-n+\ell+1$.
Since $m'=m$ or $m'=m+1$, and also $\Lambda^{Q}_{n}=0$ for $n>m'$
and $\Lambda^{P}_{n}=0$ for $n>m$, we may extend the intervals of
summation to $0\le\ell,n\le m'$, (also noting
$\sum_{i=k}^la_i\equiv-\sum_{i=l+1}^{k-1}a_i$ for $l<k$),
so that
\begin{equation}
\fl{\mbox{\myeu h}}^{Q,P}(z'_k,z)=
\sum_{\ell=0}^{m'}\sum_{n=0}^{m'}\sum_{i=\ell+1}^{m+\ell-n}
\bigg(\frac {z'_{k}}z\bigg)^\ell z_{\vp}^{i+n-1}
\big[(i-\ell)\Lambda^{Q}_{n}\Lambda^{P}_{i}\!-
(n\!-\!\ell)\Lambda^{Q}_{i}\Lambda^{P}_{n}\big].
\label{hqp3}\end{equation}
To evaluate this we enlarge the summation interval for $i$ from
$\ell+1\le i\le m+\ell-n$ to $0\le i\le m'$ and subtract the
contributions of $0\le i\le \ell$ and $m+\ell-n+1\le i\le m'$.
More precisely, defining
\begin{equation}
\alpha(z)=\sum_{n=0}^{m'}\sum_{i=0}^{m'}\,
[(i-\ell)\Lambda^{Q}_{n} \Lambda^{P}_{i}
-(n-\ell) \Lambda^{Q}_{i}\Lambda^{P}_{n}]z^{i+n-1},
\label{alpha}\end{equation}
\begin{equation}
\beta(z)=\sum_{n=0}^{m'}\sum_{i=0}^{\ell}\,
[(i-\ell)\Lambda^{Q}_{n} \Lambda^{P}_{i}
-(n-\ell) \Lambda^{Q}_{i}\Lambda^{P}_{n}]z^{i+n-1},
\label{beta}\end{equation}
and
\begin{equation}
\gamma(z)=\sum_{n=0}^{m'}\,\sum_{i=m+\ell-n+1}^{m'}\,
[(i-\ell)\Lambda^{Q}_{n} \Lambda^{P}_{i}
-(n-\ell) \Lambda^{Q}_{i}\Lambda^{P}_{n}]z^{i+n-1}.
\label{gamma}\end{equation}
we have
\begin{eqnarray}
{\mbox{\myeu h}}^{Q,P}(z'_k,z)=\sum_{\ell=0}^{m'}
\bigg(\frac {z'_{k}}z\bigg)^\ell\big[\alpha(z)-\beta(z)-\gamma(z)\big].
\label{hk2}\end{eqnarray}

It is easily seen that $\alpha(z)=0$ by interchanging
$n\leftrightarrow i$ in the sum of the second term of the summand.
Similarly, as in \cite{AuYangPerk2010a}, we can show that
for $n\le\ell$, so that $i\ge m+1,m'$, the sum over $i$ vanishes.
This leaves
\begin{eqnarray}
\gamma(z)=\sum_{n=\ell+1}^{m'}\,\sum_{i=m+\ell-n+1}^{m'}
[(i-\ell)\Lambda^{Q}_{n} \Lambda^{P}_{i}
-(n-\ell) \Lambda^{Q}_{i}\Lambda^{P}_{n}]z^{i+n-1}=0,
\label{gamma1}\end{eqnarray}
which is identically zero when we interchange the order of the
summations over $i$ and $n$ for the first term in the summand
(while noting again $\Lambda^P_{m'}=0$ if $m'=m+1$) and
make the interchange $n\leftrightarrow i$ for the second term.

Consequently, the only nonvanishing term is
\begin{eqnarray}
{\mbox{\myeu h}}^{Q,P}(z'_k,z)
=-\sum_{\ell=0}^{m'}{\left(\frac{z'_{k}}z\right)}^\ell\beta(z)
\nonumber\\
=\sum_{n=0}^{m'}\sum_{i=0}^{m'}z^{n+i-1}
\bigg[(n\Lambda^{Q}_{i} \Lambda^{P}_{n}
-i \Lambda^{Q}_{n}\Lambda^{P}_{i})
\sum_{\ell=i}^{m'}{\left(\frac{z'_{k}}z\right)}^\ell
\nonumber\\
\qquad
-(\Lambda^{Q}_{i} \Lambda^{P}_{n}
-\Lambda^{Q}_{n}\Lambda^{P}_{i})
\sum_{\ell=i}^{m'}\ell{\left(\frac{z'_{k}}z\right)}^\ell\bigg].
\label{hk3}\end{eqnarray}
Using
\begin{equation}
\sum_{\ell=i}^{m'}u^\ell=\frac{u^{m'+1}-u^i}{u-1},\qquad
\sum_{\ell=i}^{m'}\ell u^{\ell}=u\frac d{du}\sum_{\ell=i}^{m'}u^\ell,
\label{add}\end{equation}
and
\begin{equation}
\sum_{j=0}^{m}\Lambda^{P}_ju^j=P_P(u),\quad
\sum_{j=0}^{m'}j\Lambda^{Q}_ju^{j-1}
=P'_Q(u),\quad P_Q(z'_{k})=0,
\label{DP}\end{equation}
we find
\begin{equation}
{\mbox{\myeu h}}^{Q,P}(z'_k,z)
=\frac {z'_{k}P_P(z)}{(z'_k-z)}P'_Q(z'_{k})+
\frac {z'_{k}P_Q(z)}{(z'_k-z)^2}P_P(z'_{k}).
\label{hk4}\end{equation}
Substituting (\ref{dh}) and (\ref{hk4}) into (\ref{oeeo}), we find
\begin{eqnarray}
\fl\langle\Omega|{\bf E}_{j,Q}^{-}{\bf E}_{\ell,P}^{+}|\Omega\rangle=
-\frac{\beta^Q_{j,0}\beta^P_{\ell,0}z\vp_{\ell}}
{\Lambda^Q_{0}\Lambda^P_{0}}{\mbox{\myeu h}}^{Q,P}(z'_j,z\vp_\ell)=
-\frac{\beta'_{j,0}\beta\vp_{\ell,0}z\vp_{\ell}z'_{j}
P_P(z'_{j})P_Q(z\vp_{\ell})}
{\Lambda'_{0}\Lambda\vp_0(z'_j-z\vp_\ell)^2}.
\label{oeeo2}
\end{eqnarray}
From (III.17) and (IV.68) we have
\begin{eqnarray}
P_P(z)=\Lambda_m\prod_{k=1}^{m}(z-z_k),\quad
\beta_{\ell,0}=-\frac{\Lambda_0}{\Lambda_m z_\ell}
\prod_{k=1,k\ne \ell}^{m}\frac 1{(z_\ell-z_k)},
\nonumber\\
P_Q(z)=\Lambda'_m\prod_{k=1}^{m'}(z-z'_k),\quad
\beta'_{j,0}=-\frac{\Lambda'_0}{\Lambda'_{m'}z'_j}
\prod_{k=1,k\ne j}^{m'}\frac 1{(z'_j-z'_k)}.
\label{PB}\end{eqnarray}
Substituting these into (\ref{oeeo2}) we obtain
\begin{eqnarray}
\langle\Omega|{\bf E}_{j,Q}^{-}{\bf E}_{\ell,P}^{+}|\Omega\rangle=
\prod_{k=1,k\ne j}^{m'}\frac{(z\vp_\ell-z'_k)}{(z'_j-z'_k)}
\prod_{k=1k\ne \ell}^{m}\frac{(z'_j-z\vp_k)}{(z\vp_\ell-z\vp_k)}.
\label{oeeo3}\end{eqnarray}
This is exactly the form in (\ref{nEE}) with $W\vp_1\!=\!\{\ell\}$
and $W'_1\!=\!\{j\}$.

Because of the symmetry (\ref{hpqhqp}), we find from (\ref{oeeo}) that
\begin{eqnarray}
\langle\Omega|{\bf E}_{\ell,P}^{-}{\bf E}_{j,Q}^{+}
|\Omega\rangle=-({\beta^Q_{j,0}
\beta^P_{\ell,0}z'_{j}}/{\Lambda^Q_{0}\Lambda^P_{0}})
{\mbox{\myeu h}}^{P,Q}(z\vp_\ell,z'_j)
\nonumber\\
=-({\beta^Q_{j,0}\beta^P_{\ell,0}z'_{j}}/{\Lambda^Q_{0}\Lambda^P_{0}})
{\mbox{\myeu h}}^{Q,P}(z'_j,z\vp_\ell)=({z'_{j}}/{z\vp_\ell})
\langle \Omega|{\bf E}_{j,Q}^{-}{\bf E}_{\ell,P}^{+}|\Omega\rangle.
\label{oeeo3a}\end{eqnarray}

\subsection{Proof by induction\label{sec3.2}}
Now we prove the Proposition by induction. The idea of the proof
is inspired by reading paper \cite{BaxterV} by Baxter. Denote
\begin{eqnarray}
\psi_n(W\vp_n,W'_n)=\langle\Omega|\prod_{j\in W_n'}{\bf E}^-_{j,Q}
\prod_{j\in W\vp_n}{\bf E}^+_{j,P}|\Omega\rangle,
\nonumber\\
{\bar\psi}\vp_n(W\vp_n,W'_n)=\langle\Omega|
\prod_{i\in W\vp_n}{\bf E}^-_{i,P}
\prod_{j\in W'_n}{\bf E}^+_{j,Q}|\Omega\rangle.
\label{fn}\end{eqnarray}
We have so far shown that for $n=1$, the following holds:
\begin{eqnarray}
\psi\vp_n(W\vp_n,W'_n)=\frac {\bA_{s,s'}\bB_{s,s'}}{\bC_{s}\bD_{s'}},
\label{psin}\\
{\bar\psi}\vp_n(W\vp_n,W'_n)=\psi\vp_n(W\vp_n,W'_n)
\prod_{i\in W\vp_n}z^{-1}_i\prod_{j\in W'_n}z'_j\,,
\label{barpsin}\end{eqnarray}
where $\bA$, $\bB$, $\bC$ and $\bD$ are given in (\ref{ABCD}).
Substituting these we can summarize (\ref{fn}) as
\begin{equation}
\psi\vp_n(W\vp_n,W'_n)=
\frac{\displaystyle{\prod_{i\in W\vp_n}\prod_{j\in V'_n}(z\vp_i-z'_j)
\prod_{i\in W'_n}\prod_{j\in V\vp_n}(z'_i-z\vp_j)}}
{\displaystyle{\prod_{i\in W\vp_n}\prod_{j\in V\vp_n}(z\vp_i-z\vp_j)
\prod_{i\in W'_n}\prod_{j\in V'_n}(z'_i-z'_j)}}\;
\frac{\displaystyle{\prod_{i\in W\vp_n}\phi(z\vp_k)}}
{\displaystyle{\prod_{j\in W'_n}\phi(z'_\ell)}},
\label{allpsin}\end{equation}
choosing $\phi(z)=1$ for $\psi\vp_n(W\vp_n,W'_n)$ and
$\phi(z)=1/z$ for $\bar\psi\vp_n(W\vp_n,W'_n)$.

Assuming (\ref{nEE}), or more explicitly (\ref{allpsin}), holds up
to $n$, we shall prove that it also holds for $n+1$.

\subsection{Proof for $n+1$\label{sec3.3}}
Consider
\begin{eqnarray}
\psi\vp_{n+1}(W\vp_{n+1},W'_{n+1})=
\langle\Omega|{\raisebox{-4pt}{$\Bigg[$}}
\prod_{j\in W_n'}{\bf E}^-_{j,Q}{\raisebox{-4pt}{$\Bigg]$}}
{\bf E}^-_{\ell,Q}{\bf E}^+_{k,P}{\raisebox{-4pt}{$\Bigg[$}}
\prod_{j\in W_n}{\bf E}^+_{j,P}{\raisebox{-4pt}{$\Bigg]$}}
|\Omega\rangle,
\label{fn+1}\end{eqnarray}
with $W\vp_{n+1}\!=\!\{W\vp_n,k\}$ and $W'_{n+1}\!=\!\{W'_n,\ell\}$,
so that $V\vp_{n+1}\!=\!V\vp_n/\{k\}$ and
$V'_{n+1}\!=\!V'_n/\{\ell\}$, where $V/U\!=\!\{j\in V,\,j\notin U\}$.
The operators ${\bf E}^+_{j,Q}$ and ${\bf E}^-_{j,Q}$ were defined
for $Q=0$ in \cite{AuYangPerk2009} and for general $Q$ in
\cite{AuYangPerk2010b}. For fixed $Q$ they satisfy the commutation
relations in (II.15) and generate a direct sum of spin 1/2
representations of $\mathfrak{sl}_2$ algebras, see also the
appendix of \cite{AuYangPerk2009}.

Because this implies $({\bf E}^+_{j,P})^2\!=\!0$ and
$({\bf E}^-_{j,Q})^2\!=\!0$, we find that
\begin{eqnarray}
\psi\vp_{n+1}(W\vp_{n+1},W'_{n+1})=0\quad
\hbox{if}\quad \ell\in W'_n\quad \hbox{or}\quad k\in W\vp_n.
\label{cfn+1}\end{eqnarray}
From (\ref{eo}), we see that to each of the roots $z'_\ell$ of
the polynomial $P_Q(z)$, we associate an operator ${\bf E}^-_{\ell,Q}$.
Thus $\ell\in W'_n$ is equivalent to $z'_\ell= z'_j$ for some
$j\in W'_{n}$. Similarly, $k\in W_n$ corresponds to $z_k\!=\!z_j$
for some $j\in W_{n}$.

If we let ${\bf E}^-_{\ell,Q}\to{\bf E}^-_{i,P}$, (or $Q\!=\!P$,
$i\!=\!\ell$), then we must have $z'_\ell\!=\! z\vp_i$. There are three
possibilities: If $i\notin W_{n+1}$, or equivalently $i\in V_{n+1}$,
then $\psi_{n+1}(W_{n+1},W'_{n+1})\!=\!0$, due to the commutation
relation $[{\bf E}^+_{\ell,P},{\bf E}^-_{j,P}]\!=\!
\delta\vp_{\ell,j}{\bf H}\vp_{\ell,P}$. If, however
${\bf E}^-_{\ell,Q}\to{\bf E}^-_{k,P}$ making $z'_\ell\!=\! z\vp_k$ then
\begin{equation}
\fl\psi\vp_{n+1}(W\vp_{n+1},W'_{n+1})\!\to\!
\langle\Omega|{\raisebox{-4pt}{$\Bigg[$}}
\prod_{j\in W_n'}\!\!{\bf E}^-_{j,Q}{\raisebox{-4pt}{$\Bigg]$}}
{\bf E}^-_{k,P}{\bf E}^+_{k,P}{\raisebox{-4pt}{$\Bigg[$}}
\prod_{j\in W_n}\!\!{\bf E}^+_{j,P}
{\raisebox{-4pt}{$\Bigg]$}}|\Omega\rangle=\psi\vp_n(W\vp_n,W'_n),
\label{fnn}\end{equation}
as ${\bf H}\vp_{k,P}|\Omega\rangle\!=\!-|\Omega\rangle$. The third case
$i\!=\!j\in W_{n}$ is similar making $z'_\ell\!=\! z\vp_j$
leading to a reduction to $\psi\vp_n(\hat W\vp_n,W'_n)$,
with $\hat W\vp_n=W\vp_{n+1}/\{\ell\}$.

Likewise, if ${\bf E}^+_{k,P}\to{\bf E}^+_{i,Q}$, or $z\vp_k= z'_i$,
for $i\in V'_{n+1}$, then $\psi\vp_{n+1}(W\vp_{n+1},W'_{n+1})\!=\!0$.
For $ z\vp_k= z'_\ell$ we have
$\psi\vp_{n+1}(W\vp_{n+1},W'_{n+1})\!=\!\psi\vp_n(W\vp_n,W'_n)$,
whereas for $i\!=\!j\in W'_{n}$ we have another reduction to
$\psi\vp_n(W\vp_n,\hat W'_n)$, with $\hat W'_n=W'_{n+1}/\{k\}$.

To satisfy all these conditions, we must have
\begin{eqnarray}
\fl\psi\vp_{n+1}(W\vp_{n+1},W'_{n+1})=
\prod_{j\in W'_n}\frac{z'_\ell-z'_j}{z\vp_k-z'_j}
\prod_{j\in W\vp_n}\!\frac{z\vp_k-z\vp_j}{z'_\ell-z\vp_j}
\prod_{j\in V\vp_{n+1}}\frac{z'_\ell-z\vp_j}{z\vp_k-z\vp_j}
\prod_{j\in V'_{n+1}}\frac{z\vp_k-z'_j}{z'_\ell-z'_j}
\nonumber\\
\times\,\frac{\phi(z\vp_k)}{\phi(z'_\ell)}\,
\psi\vp_n(W\vp_n,W'_n),
\label{cfn+12}\end{eqnarray}
where $\phi(z)$ is a function to be determined. Note that
(\ref{cfn+12}) has all the desired zeroes and that all fractions in
it equal one for $z'_\ell\!=\!z\vp_k$. Because of the reduction to
(\ref{oeeo3}) for $n\!=\!1$, there can only be simple zeroes.
Setting $n\!=\!0$, noting that $\psi_0(\emptyset,\emptyset)=1$ is
the normalization of the state $|\Omega\rangle$, we would at first
have found the more general
\begin{equation}
\psi\vp_{1}(W\vp_{1},W'_{1})=
\prod_{j\not=k}\frac{z'_\ell-z\vp_j}{z\vp_k-z\vp_j}
\prod_{j\not=\ell}\frac{z\vp_k-z'_j}{z'_\ell-z'_j}\;
\Phi(z\vp_k,z'_\ell),\quad \Phi(z\vp_k,z\vp_k)=1,
\end{equation}
instead of (\ref{cfn+12}) for this case. But different reductions
from $\psi_n$'s with $n>1$ down to $n\!=\!0$ lead to consistency
conditions that are satisfied when
$\Phi(z\vp_k,z'_\ell)=\phi(z\vp_k)/\phi(z'_\ell)$.

So far in this subsection, the reasoning works for both $\psi_n$
and $\bar\psi_n$, so that (\ref{cfn+12}) is valid for both.
However, we still have to compare with (\ref{oeeo3}) and
(\ref{oeeo3a}). This leads to the conclusion that we must have
$\phi(z)=1$ for $\psi_n$ and $\phi(z)=1/z$ for $\bar\psi_n$.

Because (\ref{allpsin}) is assumed to hold for $n$, we substitute it
into the above equation (\ref{cfn+12}). Using
$V'_n=\{V'_{n+1},\ell\}$, $W\vp_{n+1}=\{W\vp_n,k\}$, we have
\begin{eqnarray}
\prod_{i\in W\vp_n}\prod_{j\in V'_n}(z\vp_i-z'_j)\bigg/
\prod_{i\in W\vp_n}(z\vp_i-z'_\ell)=
\prod_{i\in W\vp_n}\prod_{j\in V'_{n+1}}(z\vp_i-z'_j),
\nonumber\\
\prod_{j\in V'_{n+1}}(z\vp_k-z'_j)
\prod_{i\in W\vp_n}\prod_{j\in V'_{n+1}}(z\vp_i-z'_j)=
\prod_{i\in W\vp_{n+1}}\prod_{j\in V'_{n+1}}(z\vp_i-z'_j),
\label{cfn+13}\end{eqnarray}
and we can write six similar relations. Thus we find
\begin{equation}
\fl\psi\vp_{n+1}(W\vp_{n+1},W'_{n+1})=
\frac{\displaystyle{\prod_{i\in W\vp_{n+1}}
\prod_{j\in V'_{n+1}}(z\vp_i-z'_j)
\prod_{i\in W'_{n+1}}\prod_{j\in V\vp_{n+1}}(z'_i-z\vp_j)}}
{\displaystyle{\prod_{i\in W\vp_{n+1}}
\prod_{j\in V\vp_{n+1}}(z\vp_i-z\vp_j)
\prod_{i\in W'_{n+1}}\prod_{j\in V'_{n+1}}(z'_i-z'_j)}}\;
\frac{\displaystyle{\prod_{k\in W\vp_{n+1}}\phi(z\vp_k)}}
{\displaystyle{\prod_{\ell\in W'_{n+1}}\phi(z'_\ell)}}.
\end{equation}
This then completes the proof by induction, establishing
(\ref{psin}) and (\ref{barpsin}).

\subsection{Inner products\label{sec3.4}}
Using (\ref{nEE}), (\ref{fn}), (\ref{barpsin}) and (\ref{ypj}),
we find (\ref{yQyP}),
(\ref{yPyQ}) become
\begin{equation}
\langle{\mathcal Y}_{\emptyset}^Q|{\mathcal Y}^{\vP}_{\emptyset}\rangle=
{\mathcal C}{\hat D}_{PQ},\quad
\langle{\mathcal Y}_{\emptyset}^{\vP}|{\mathcal Y}^Q_{\emptyset}\rangle=
{\hat {\mathcal C}}{\hat D}_{QP},\quad
{\hat D}_{QP}={\hat D}_{PQ},
\label{sum1}\end{equation}
where
\begin{equation}
{\hat D}_{PQ}=\sum_{s\vp}\sum_{s'}u_1^{s_1}u_2^{s_2}\cdots u_m^{s_m}
\frac{\bA_{s,s'}\bB_{s,s'}}{\bC_{s}\bD_{s'}}
{u'}_1^{s'_1} { u'}_2^{s'_2}\cdots{u'}_{m'}^{s'_{m'}}.
\label{DPQ}\end{equation}

\subsection{Comparison with Baxter's sum\label{sec3.5}}
Since $c=(z+1)/(z-1)$, we find
\begin{equation}
z\vp_i-z'_j=\frac{-2(c\vp_i-c'_j)}{(c\vp_i-1)(c'_j-1)},\quad
 z'_i-z\vp_j=\frac{-2(c'_i-c\vp_j)}{(c'_i-1)(c\vp_j-1)},
\end{equation}
and similar expressions for $z\vp_i-z\vp_j$ and $z'_i-z'_j$, so that
\begin{eqnarray}
\fl\bA_{s,s'}\!=\!\frac{(-2)^{n(m'-n)}A_{s,s'}}
{\displaystyle{\prod_{i\in W}(c\vp_i\!-\!1)^{m'-n}
\prod_{j\in V'}(c'_j\!-\!1)^n}},\quad
 \bB_{s,s'}\!=\!\frac{(2)^{n(m-n)}B_{s,s'}}
{\displaystyle{\prod_{i\in W'}(c'_i\!-\!1)^{m-n}
 \prod_{j\in V}(c\vp_j\!-\!1)^n}},
\nonumber\\
 \fl\bC_{s}\!=\!\frac{(2)^{n(m-n)}C_{s}}
{\displaystyle{\prod_{i\in W}(c\vp_i\!-\!1)^{m-n}
 \prod_{j\in V}(c\vp_j\!-\!1)^n}},\qquad
 \bD_{s'}\!=\!\frac{(-2)^{n(m'-n)}D_{s'}}
{\displaystyle{\prod_{i\in W'}(c'_i\!-\!1)^{m'-n}
 \prod_{j\in V'}(c'_j\!-\!1)^n}},
\end{eqnarray}
and
\begin{equation}
\frac {\bA_{s,s'} \bB_{s,s'}}{ \bC_{s}\bD_{s'}}=
\frac {A_{s,s'} B_{s,s'}}{ C_{s}D_{s'}}
\frac{{\prod_{i\in W'}(c'_i\!-\!1)^{m'-m}}}
{{\prod_{i\in W}(c_i\!-\!1)^{m'-m}}}.
\label{ABCD2}\end{equation}
For $m=m'$, the $c\vp_i-1$ and $c'_i-1$ factors in (\ref{ABCD2})
cancel out. For $m'=m+1$, we find from (\ref{lz}) and (\ref{yj4}) that
\begin{eqnarray}
c_j-1=\frac2{z_j-1}=\frac{(1-k')^2-\lambda_j^2}{2k'},
\\
\frac{u_j}{c_j-1}=\frac{-2k'}{(\lambda_j+1)^2-k'^2},\quad
u'_j(c'_j-1)=\frac{(\lambda'_j-1)^2-k'^2}{2k'}.
\end{eqnarray}
Comparing with (BaxIV.A.1), one can see that the results agree.
This shows that we have obtained the sum in (BaxIII.3.48) by a
completely different route. We can now finish the calculation of
the order parameter following Baxter \cite{BaxterIV,BaxterV} closely,
apart from some subtle differences due to different choice of
parameters. Therefore, we shall not present the details of our
calculation in the next subsection.

\subsection{Cauchy determinant\label{sec3.6}}
In the most recent paper \cite{BaxterV}, Baxter has proven that
the sum in (\ref{LHS}) is a determinant. In the same way, we may
write (\ref{DPQ}) as
\begin{equation}
{\hat D}_{PQ}=\det[\,\Isf_m+\Ysf\,\Bsf\,\Ysf'\Bsf^{\!\mathrm{T}}],
\label{sum2}\end{equation}
where we may choose to use either the variables $c\vp_i$, $c'_i$,
$y\vp_i$ and $y'_i$ of Baxter \cite{BaxterIII} or our variables
$z\vp_i$, $z'_i$, $u\vp_i$ and $u'_i$.
With the latter choice we have matrix elements
\begin{equation}
B\vp_{ij}=\frac{f\vp_if'_j}{z\vp_i-z'_j},\quad
Y\vp_{ij}=\delta\vp_{ij}u\vp_i,\quad
Y'_{ij}=\delta\vp_{ij}u'_i,
\end{equation}
with $u\vp_i$ and $u'_i$ given in (\ref{yj4}).
The constants $f\vp_i$ and $f'_i$ in this case are related to the
Drinfeld polynomials via
\begin{eqnarray}
f_i^2=\frac{\epsilon a\vp_i}{b\vp_i},\quad
{f'_i}^2=-\frac{\epsilon a'_i}{b'_i},\quad
\epsilon=\pm 1,
\nonumber\\
\fl a\vp_i=\prod_{j=1}^{m'} (z\vp_i-z'_j)=
P\vp_Q(z\vp_i)/\Lambda_{m'}^Q\,,\quad
a'_i=\prod_{j=1}^{m}(z'_i-z\vp_j)=P\vp_P(z'_i)/\Lambda_m^P\,,
\nonumber\\
\fl b\vp_i=\prod_{j=1,\;j\neq i}^m(z\vp_i-z\vp_j)=
P'_P(z\vp_i)/\Lambda_m^P\,,\quad
b'_i=\prod_{j=1,\;j\neq i}^{m'}(z'_i-z'_j)=
P'_Q(z'_i)/\Lambda_{m'}^Q\,,
\end{eqnarray}
so that $\Bsf$ is orthogonal in the sense that
\begin{equation}
\Bsf^{\!\mathrm{T}}\Bsf= \Isf_{m'} \; \; {\rm if} \; m\geq m',\qquad
\Bsf\,\Bsf^{\!\mathrm{T}} = \Isf_{m} \; \; {\rm if} \; m\leq m',
\label{orth}\end{equation}
see subsections 6.1 and 6.2 of \cite{BaxterII}. The result is not
dependent on using $c$'s or $z$'s.

Again, comparing with Baxter we have to take account of $z_j=1/w_j$,
[so that $c_j$ still lies in the interval $(-1,1)$]. For $P=0$
both $z_j$ and $1/z_j$ are roots of the polynomial $P_0(z)$.
From (\ref{eo}) we can see that to assign $1/z_j$ instead of $z_j$
to ${\bf E}_{j,0}^{\pm}$ is merely a change of convention.
For $P>0$, if $z_j$ is a root of $P_P(z)$, then $1/z_j$ is a
root of $P_{N-P}(z)$, and vice versa. The particular choice
in (\ref{eo}) is no longer arbitrary; it is chosen such that
the ${\bf E}_{j,P}^{\pm}$ are to act on the ground state sector
of spin-translation quantum number $P$. Due to this complication,
there are some minus sign differences presented below.

Instead of (BaxIV.4.21), we find for $m=m'$,
\begin{eqnarray}
{\hat D}_{PQ}=\frac{\Delta_{m,m}(\lambda,\lambda')}
{\Delta_{m,m}(\lambda^2,\lambda'{}^2)}
\prod_{i=1}^m\frac 2{(1+k'+\lambda\vp_i)(1-k'+\lambda'_i)},
\label{hdpq}\end{eqnarray}
where, as in (BaxIV.2.8),
\begin{equation}
\Delta_{m,m'}(c,c')=
\frac{\displaystyle{\prod_{1 \leq i< j\leq m}(c\vp_i-c\vp_j)
\prod_{1\leq i< j\leq{m'}}(c'_j-c'_i)}}
{\displaystyle{\prod_{i=1}^m\prod_{j=1}^{m'}(c\vp_i-c'_j)}}.
\label{defDelta}\end{equation}
Therefore, using (\ref{ss}), (\ref{ypj}) and (\ref{sum1}),
we find for $m=m'$
\begin{eqnarray}
\langle{\mathcal Y}_{\emptyset}^Q|{\mathcal Y}^{\vP}_{\emptyset}\rangle
\langle{\mathcal Y}_{\emptyset}^{\vP}|{\mathcal Y}^Q_{\emptyset}\rangle=
\frac{{\mathcal R}(1-k')}{{\mathcal R}(1+k')}
\prod_{i=1}^m\frac{{\mathcal R}(\lambda'_i)}{{\mathcal R}(\lambda\vp_i)},
\label{formf}\end{eqnarray}
where, as in (BaxIV.3.8), we define
\begin{equation}
{\mathcal R}(\lambda)=\frac{\prod_{i=1}^{m}
\,(\lambda+\lambda\vp_i)/2}
{\prod_{j=1}^{m'}\,(\lambda+\lambda'_j)/2}\,.
\end{equation}
However, because $z_j=1/w_j$, instead of (BaxIV.5.8) and
(BaxIV.5.12), we have here
\begin{equation}
{\mathcal R}(\lambda)=\left(\frac{1-k'+\lambda}{2}\right)^{m-m'}
\,\left(\frac{1+k'+\lambda}{1-k'+\lambda}\right)^{(Q-P)/N},
\end{equation}
\begin{equation}
\fl{\mathcal R}(1-k')=(1-k')^{m-m'+(P-Q)/N},\quad
{\mathcal R}(1+k')=(1+k')^{(Q-P)/N}.
\label{calcR}\end{equation}
As a consequence, we have from (\ref{formf}) the spontaneous
magnetization given as
\begin{eqnarray}
\fl\left({\mathcal M}_r\right)^2=
\langle{\mathcal Y}_{\emptyset}^Q|{\mathcal Y}^{\vP}_{\emptyset}\rangle
\langle{\mathcal Y}_{\emptyset}^{\vP}|{\mathcal Y}^Q_{\emptyset}\rangle=
(1-k'{}^2)^{(P-Q)(N-P+Q)/N^2}
\nonumber\\
=(1-k'{}^2)^{r(N-r)/N^2}.
\label{order2}\end{eqnarray}

For $m'=m+1$, we find the same result as (BaxIV.A.17), namely 
\begin{equation}
{\hat D}_{PQ}=\frac{\Delta_{m,m+1}(\lambda,\lambda')}
{\Delta_{m,m+1}(\lambda^2,\lambda'{}^2)}
\prod_{i=1}^m\frac 2{(1+\lambda_i)^2-k'{}^2},
\label{hdpq2}\end{equation}
so that
\begin{eqnarray}
\left({\mathcal M}_r\right)^2=
\langle{\mathcal Y}_{\emptyset}^Q|{\mathcal Y}^{\vP}_{\emptyset}\rangle
\langle{\mathcal Y}_{\emptyset}^{\vP}|{\mathcal Y}^Q_{\emptyset}\rangle=
{\mathcal C}{\hat {\mathcal C}} {\hat D}_{PQ}^{\;2}
\nonumber\\
=\frac 1{{\mathcal R}(1-k'){\mathcal R}(1+k')}\,
\frac{\prod_{i=1}^{m'}{\mathcal R}(\lambda'_i)}
{\prod_{i=1}^{m}{\mathcal R}(\lambda\vp_i)}=(1-k'{}^2)^{r(N-r)/N^2},
\label{order3}\end{eqnarray}
which is the same as (\ref{order2}). Thus from (\ref{order2}) and
(\ref{order3}), we see that the $N$ terms in the sum given in
(\ref{order}) are all equal in the thermodynamic limit.

\section{Summary and Outlook\label{sec4}}
In this paper we have discussed two approaches to the pair correlation
function in the chiral Potts model. In subsection \ref{sec1.2} we noted
that the ground state eigenvectors of \cite{AuYangPerk2009} suffice
for the calculation of the pair correlation in the superintegrable
chiral Potts chain in the commensurate phase. In fact, to calculate its
correlation $\langle{\bf{\hat Z}}_1^r{\bf{\hat Z}}_{R+1}^{-r}\rangle$,
we need to evaluate
\begin{equation}
\langle\Omega|\prod_{k=1}^n {\bf E}^-_{j_k,P}
({\bf Z}^r_1\cdots{\bf Z}^r_R)
\prod_{k=1}^n {\bf E}^+_{\ell_k,P}|\Omega\rangle,
\end{equation}
which we do not know how to evaluate yet for arbitrarily large system
size. However, for $N=3$ and $L=3$, it yields a result identical to that
of Fabricius and McCoy \cite{FabMcC2010}. In subsection \ref{sec1.1}
we explained that the knowledge of the transfer matrix eigenvectors
of the superintegrable chiral Potts should suffice for the calculation
of the pair correlation in the more general integrable chiral Potts
model. 

In our previous papers \cite{AuYangPerk2009,AuYangPerk2010a} we have
determined $2^{m_Q}$ eigenvectors for each $Q$ sector and they are
given here in (\ref{eigenvectors}). We have learned to calculate
inner products of such states chosen from different $Q$ sectors by
evaluating the inner product of the state with $n$ creation operators
${\bf E}^+_{j,P}$ acting on $|\Omega\rangle$ with one with $n$
annihilation operators ${\bf E}^-_{j,Q}$ acting on $\langle\Omega|$
given in (\ref{nEE}). We have used these inner products to calculate
the pair correlation function in the large separation limit, which
gave us the order parameter. This derivation uses (\ref{Id}), which
in earlier preprint versions of this paper had to be left as a
conjecture, but finally got proved \cite{AuYangPerk2011} in a
rather lengthy manner.

However, if we want to go further and study the separation dependence
of the pair correlation function using (\ref{gr2l}), we will need to
know the inner products for all eigenvectors $|{\mathcal Y}_j^Q\rangle$
that are not orthogonal to the maximum eigenvectors. One way is to
first calculate all eigenvectors of the superintegrable $\tau_2$ model
and the related higher-spin XXZ-model by Bethe Ansatz working out
the prescription of Tarasov \cite{Tarasov,Roan,Roan2}, and then to
make proper linear combinations in each degenerate eigenvalue space
\cite{Roan2} as we have done \cite{AuYangPerk2009,AuYangPerk2010a} for
the ground state sector using Onsager algebra and quantum loop algebra.
But it will be very hard to work out all necessary details this way.
We believe that there should be a better way using more of the
combinatorial structure of the problem, which we may start exploring
using the approach proposed in subsection \ref{sec1.2}, for which we
have the needed eigenvectors explicitly.

\section*{Acknowledgments}

We thank Prof.\ Rodney Baxter for sharing valuable notes and helpful
communications. This work was first presented at the Simons Center
for Geometry and Physics Workshop on Correlation Functions for
Integrable Models 2010: January 18--22, 2010. We thank Prof.\ Barry
McCoy and the Simons Foundation for their kind invitation.
\appendix
\section{Generating Function ${\mathcal G}(t,u)$\label{secA}}
The generating functions $g$ and $\bar g$ are defined in (II.62),
(III.14) and (III.15) as
\begin{eqnarray}
g(\{n_j\},t)=\sum_{m=0}^{(N-1)L- N}K_m(\{n_j\})t^m,
\nonumber\\
{\bar g}(\{n_j\},t)=\sum_{m=0}^{(N-1)L-N}\bK_m(\{n_j\})t^m.
\label{gt}\end{eqnarray}
It was shown in \cite {AuYangPerk2010a} that they have
the simple form
\begin{equation}
g(\{n_j\},t)={\bar g}^*(\{n_j\},t)
=(1-t^N)^{L-1}\prod_{j=1}^L\,(1-t\omega^{N_j})\strut^{-1}.
\label{gtf}\end{equation}
We now define the two-variable generating function
\begin{equation}
{\mathcal G}(t,u)=\sum_{ {\{0\le n_j\le N-1\}}\atop{n_1+\cdots+n_L=N} }
{\bar g}(\{n_j\},t)\,g(\{n_j\},u).
\label{Gtu}\end{equation}
Substituting (\ref{gt}) into the above equation and comparing with
(\ref{coeff}), we find
\begin{equation}
{\mathcal G}(t,u)=\sum_{\ell=0}^{N(r-1)}\sum_{k=0}^{N(r-1)}
{\mathcal G}_{\ell,k} t^\ell u^k, \qquad r=(N-1)L/N.
\label{gtu}\end{equation}

From Baxter's paper \cite{BaxterIII}, we realized that the
coefficients ${\mathcal G}_{\ell,k}$ must be related to the
coefficients of the Drinfeld polynomials (\ref{dfp}). However,
the method that we used in \cite{AuYangPerk2010a} cannot be used,
as is shown next.
\subsection{Difficulties for $P\ne Q$\label{secA.2}}
From (\ref{coeff}) and (\ref{KbK}), we may write
\begin{equation}
\fl{\mathcal G}_{\ell N+Q,\,jN+P}
=\underset{{\{0\le\lambda\vp_i,n\vp_i,n'_i\le N-1\}}
\atop{\sum n\vp_i=N,\;\sum n'_i=\ell N+Q,\;\sum\lambda\vp_i=jN+P}}
{\sum\sum\sum}
\prod_{i=1}^L\sfactor{n'_i+n\vp_i}{n\vp_i}
\sfactor{n_i+\lambda_i}{n_i}\omega^{n\vp_i(N'_i+{\bar b}\vp_i)},
\end{equation}
where
${\bar b}\vp_j=\sum_{i>j}\lambda\vp_i$ and
$N'_j=\sum_{i<j}n'_i$.
Let $\mu\vp_i=n\vp_i+n'_i$, so that $\sum\mu\vp_i=(\ell+1)N+Q$.
Then we find
\begin{eqnarray}
\fl{\mathcal G}_{\ell N+Q,jN+P}=
\underset{{{\{0\le\lambda_i,n_i,\mu_i\le N-1\}}\atop{\sum n_i=N,\;
\sum\lambda_i=jN+P,}}\atop{\sum\mu\vp_i=(\ell+1)N+Q}}{\sum\sum\sum}
\prod_{i=1}^L\sfactor{\mu_i}{n_i}
\sfactor{n_i+\lambda_i}{n_i}\omega^{n_i(a_i-N_i+{\bar b}_i)}
\nonumber\\
=\underset{{\{0\le\mu_i,\lambda_i\le N-1\}}\atop{\sum\mu_i=(\ell+1)N+Q,
\;\sum\lambda_i=jN+P}}{\sum\sum}
{\mathcal I}_{N}(\{\mu_i\};\{\lambda_i\}),
\end{eqnarray}
where $a_j=\sum_{i<j}\mu_i$ and $N_j=\sum_{i<j}n_i$ and
\begin{equation}
{\mathcal I}_n(\{\mu_i\};\{\lambda_i\})\equiv
\sum_{{\{0\le n_i\le N-1\}}\atop{n_1+\cdots+n_L=n}}
\prod_{i=1}^L\sfactor{\mu_i}{n_i}\sfactor{n_i+\lambda_i}{n_i}
\omega^{n_i(a_i-N_i)+n_i{\bar b}_i},
\label{In}\end{equation}
see (III.22).
Using $a_{L+1}=\sum\mu_i=(\ell+1)N+Q$ and
${\bar b}_0=\sum\lambda_i=jN+P$, we find from (III.28),
(III.33), (III.34) and (III.35) the relation
\begin{eqnarray}
\fl\sum_{n=0}^{a_{L+1}}
(-1)^n\omega^{\frac12n^2}{\mathcal I}_n(\{\mu_i\};\{\lambda_i\})t^n
\nonumber\\
=(\omega^{\halfs+P}t;\omega)^{\vphantom{P}}_{N-P+Q}
(1+t^N)^{\ell-j}
\sum_{n=0}^{{\bar b}_{0}}(-1)^n\omega^{\frac12n^2}
\bI_n(\{\lambda_i\};\{\mu_i\})t^n,
\label{IbI}\end{eqnarray}
where
\begin{equation}
\bI_n(\{\lambda_i\};\{\mu_i\})\equiv
\sum_{ {\{0\le n_i\le N-1\}}\atop{n_1+\cdots+n_L=n}}
\prod_{i=1}^L\sfactor{\lambda_i}{n_i}\sfactor{n_i+\mu_i}{n_i}
\omega^{n_i({\bar b}_i-{\bar N}_i)+n_i a_i},
\label{bIn}\end{equation}
as in (III.26). By equating the coefficients of $t^n$ on both
sides of (\ref{IbI}), we relate the sums ${\mathcal I}_n$ and $\bI_n$.
For $P=Q$, this process simplifies as 
$(\omega^{\halfs+P}t;\omega)^{\vphantom{P}}_N=1+t^N$ in (\ref{IbI}).
For $P\ne Q$, it is more complicated, but we still can use
(\ref{IbI}) to relate ${\mathcal I}_N$ with $\bI_N$ and other $\bI_n$'s
with the $n<N$ coefficients calculated from (III.11).
Thus we obtain
\begin{equation}
{\mathcal G}\vp_{\ell N+Q,jN+P}
=(\ell-j)\Lambda_{\ell+1}^Q\Lambda_{j}^P+
{\mathcal G}\vp_{(\ell+1)N+Q,(j-1)N+P}+{\mathcal U}^{Q,P}_{\ell,j},
\label{GU}\end{equation}
where
\begin{equation}
\fl{\mathcal U}^{Q,P}_{\ell,j}=\sum_{k=0}^Q
\sfactor{N-P+Q}{Q-k}\omega^{k^2-kP}
\underset{{\{0\le \mu'_i,\lambda'_i\le N-1\}}\atop
{{\sum\mu'_i=(\ell+1)N+Q+P-k}\atop{\sum\lambda'_i=jN+k}}}{\sum\sum}
{\mathcal I}_{P-k}(\{\mu'_i\};\{\lambda'_i\}).
\label{UQP}\end{equation}
We have changed the $\bI_n$'s to ${\mathcal I}_n$'s by changing the
summation variables to $\lambda\vp_i=\lambda'_i+n\vp_i$ and
$\mu\vp_i=\mu'_i-n\vp_i$, where the $n\vp_i$'s are the
summation variables in (\ref{In}) and (\ref{bIn}).

For $P=Q$, the only nonzero term is $k=Q=P$. Since
${\mathcal I}_{0}(\{\mu'_i\};\{\lambda'_i\})=1$, we find
\begin{equation}
{\mathcal U}^{Q,Q}_{\ell,j}=
\sum_{\{0\le \mu'_i\le N-1\}\atop{\sum\mu_i=(\ell+1)N+Q}}
\sum_{\{0\le\lambda'_i\le N-1\}\atop{\sum\lambda_i=jN+Q}}1
=\Lambda_{\ell+1}^Q\Lambda_{j}^Q,
\label{UQQ}\end{equation}
so that (\ref{GU}) becomes identical to (III.43) as it should be.

For $P>Q$, we have to keep using (\ref{IbI}) to express each
${\mathcal I}_{k}(\{\mu'_i\};\{\lambda'_i\})$ in (\ref{UQP}) in terms
of ${\mathcal I}_{\ell}(\{\mu_i\};\{\lambda_i\})$ with $\ell\le k$
and $n=\sum_i\lambda_i\le\sum_i\lambda'_i$, as we have done
once before to arrive at (\ref{GU}), continuing until $n\le k$,
such that ${\mathcal I}_{k}=0$. Let
\begin{equation}
\sum_{\{0\le\lambda_i\le N-1\}\atop{\sum\lambda_i=m}}1=c_m,
\quad c_{nN+Q}=\Lambda_n^Q.
\end{equation}
Since ${\mathcal I}_{0}=1$, we find it is possible to express
${\mathcal U}^{Q,P}_{\ell,j}$ as a sum of $c_{(\ell+1)N+Q+P-m}c_{jN-m}$.
However, this is very tedious and messy. Using Maple we have
discovered (and checked for different $P$, $Q$ and $N$ values
and small system sizes $L$) that the following equation holds:
\begin{equation}
{\mathcal U}^{Q,P}_{\ell,j}=
\sum_{n=0}^j\Lambda_{n}^Q \Lambda_{\ell+1+j-n}^P-
\sum_{n=0}^{j-1}\Lambda_{n}^P\Lambda_{\ell+1+j-n}^Q,
\label{UQP2}\end{equation}
so that
\begin{eqnarray}
\fl{\mathcal G}_{\ell N+Q,jN+P}
=\sum_{n=0}^j(\ell-j+2n)\Lambda_{\ell+1+n}^Q\Lambda_{j-n}^P+
\sum_{n=0}^j{\mathcal U}^{Q,P}_{\ell+n,j-n}
\nonumber\\
\fl=\sum_{n=0}^j\bigg[(\ell-j+2n)\Lambda_{\ell+1+n}^Q\Lambda_{j-n}^P+
\sum_{m=0}^{j-n}\Lambda_{m}^Q\Lambda_{\ell+1+j-m}^P
-\sum_{m=0}^{j-n-1}\Lambda_{\ell+1+j-m}^Q\Lambda_m^P\bigg]
\nonumber\\
\fl=\sum_{m=0}^j\bigg[(\ell+j-2m)\Lambda_{\ell+1+j-m}^Q\Lambda_{m}^P
+(j-m+1)\Lambda_{m}^Q\Lambda_{\ell+1+j-m}^P-(j-m)
\Lambda_{\ell+1+j-m}^Q\Lambda_m^P\bigg]
\nonumber\\
\fl=\sum_{m=0}^j\bigg[(\ell-m)\Lambda_{\ell+1+j-m}^Q\Lambda_{m}^P
+(j-m+1)\Lambda_{m}^Q\Lambda_{\ell+1+j-m}^P\bigg],
\label{GU1}\end{eqnarray}
which is the identity in (\ref{Id}). In the first preprint version
\cite{AuYangPerk2010c} of this work (\ref{UQP2}) had to be left
as a conjecture. In the meantime, a lengthy proof of (\ref{GU1})
has been found, which is presented elsewhere \cite{AuYangPerk2011}.


\end{document}